\documentclass[aps,%jmp, %letter,
scriptaddress,spanish%,%prl
] %,twocolumn]
{revtex4} 
\usepackage{amsmath, amssymb}
\usepackage{makeidx}
\usepackage{amsfonts}
\usepackage{ucs}
\usepackage[utf8]{inputenc}
\usepackage[usenames,dvipsnames]{pstricks}
\usepackage{subfigure}  % use for side-by-side figures
\usepackage{epsfig}
\usepackage{pst-grad} % For gradients
\usepackage{pst-plot} % For axes
\usepackage[colorlinks,hyperindex]{hyperref}
\usepackage{multirow}   %%% for tables multirow

\begin{document}
%\providecommand{\keywords}[1]
%{
%  \small	
%  {\textit{Keywords:~Multifractal, Hölder regularity, Isospectral, Symmetry, Riccati equation.}}
%}

\title{Multifractal analysis of the symmetry of a strictly isospectral energy landscape on a square lattice}

\author{J. de la Cruz$^{a}$, J.~S. Murguía$^b$, H.~C. Rosu$^a$}
%\date{%
    \affiliation{$^{a}$%Divisi\'on de Materiales Avanzados,
    IPICyT - Instituto Potosino de Investigación Científica y Tecnológica, \\Camino a la Presa San José 2055, Lomas $4^a$ sección, 78216, San Luis Potosí, S.L.P., Mexico\\
    $^b$Facultad de Ciencias, Universidad Autónoma de San Luis Potosí, \\Avenida Parque Chapultepec 1570, 78210, San Luis Potosí, S.L.P., Mexico}%
%}
%}

%\twocolumn[
%\begin{@twocolumnfalse}
%\maketitle
%\centerline{Materials Letters, May/June 2021}
\begin{abstract}
We use the Hölder regularity analysis to study the symmetry breaking and recovery due to a parametric potential generated via the strictly isospectral factorization method. The initial potential is two-dimensional and periodic in the two Cartesian directions,
with the symmetry group $P_{4mm}$. The resulting parametric isospectral potential display a $P_m$ symmetry for values of the parameter moderately close to the singular value $\gamma_s$. However, at large values of the parameter, visually around $\gamma=\gamma_s+110$, the original symmetry is recovered. For a much higher precision value of the parameter for this symmetry recovery,
we show that the multifractal spectrum of the parametric potential can be conveniently used. In the latter case, we obtain $\gamma=\gamma_s+201.085$ for three decimal digits precision.\\ \\
{\scriptsize \textit{Keywords:~Multifractal, Hölder regularity, Isospectral, Symmetry, Riccati equation.}} \hfill {\tiny Materials Letters, 2021}
 \end{abstract}
 %\keywords{}
%\end{@twocolumnfalse}
%]
\maketitle

\section{Introduction}

 In 1984, Mielnik \cite{Mielnik} obtained the parametric Darboux isospectral potential $V(x,\gamma)=x^2/2-d/dx\big[\log(e^{-x^2/2}+\gamma)\big]$ for the case of the harmonic oscillator potential, $V_{\rm harm}=x^2/2$, and since then many parametric isospectral potentials have been discussed in the literature \cite{rmc}. Here, Mielnik's method is applied to a periodic two-dimensional potential with identical sinusoidal components on both Cartesian axes which is of interest in textures of energy landscapes. The main goal is to obtain a very precise value (much beyond the visual one) of the Darboux deformation parameter $\gamma$ at which the symmetry associated to the original potential is recovered.
 The feature of symmetry recovery in this case is based on the well known fact that when $\gamma \rightarrow \infty$ the parametric deformation goes to zero.
From the strict calculus point of view this comes out from the fact that in equation (\ref{pi}) below, $\gamma$ stands in the denominator of $\phi$ and as such the $\phi$ term goes to zero at increasing $\gamma$. Physically, it is well known that $\gamma$ is related to the change of boundary conditions from Dirichlet to Robin ones \cite{cz14}, and so the recovery of the initial symmetry can be controlled through external fields. We notice that the issue of (super)symmetry breaking and recovery has been discussed previously for relativistic quantum field models, but focused on the concept of mass generation, see e.g. \cite{Sako-Suzuki, Bevilaqua}, while in condensed matter physics, the same issue has been related to phase transitions, either of equilibrium or non-equilibrium type.

The symmetry operations for the chosen periodic potential
  are $\pi/2$ rotations applied at the center of the cell and reflections over the axis and diagonals, while the deformed potential landscapes only possess a reflection with respect to the principal diagonal. When the deformation parameter takes a sufficiently high value the symmetry operations for the deformed potential are very close to the non deformed ones with a high level of precision and one can say that all the lost symmetries are recovered.

The effect of the deformation parameter is to modify the local maxima and minima distribution in the deformed potential, which physically implies that the probability density is redistributed over the landscape as a function of the deformation parameter, and becomes the same as that of the original potential when the deformation parameter is increased.
In practice, the level of the achieved symmetry recovery is high already at moderate values of $\gamma$, but we are interested in much more precise values of this parameter such that the initial potential landscape and the corresponding deformed one can be considered as indistinguishable.

 To find such precise values, we use the multifractal analysis which allows to study in detail the distribution of singularities of a function. The spectrum of singularities depends on the Hölder exponents that are used generically in the signal processing area since they allow the characterization of the local regularity of a signal or a function \cite{He}.  %\cite{He,Struzik,RM}.
Moreover, our multifractal approach to isospectral potential energy landscapes and their symmetry recovery displays similarities with Kolmogorov's statistical theory of turbulence based on the concepts of energy cascades and self-similarity, in particular with a symmetry model of multifractal cascades in fully developed turbulence which incorporates intermittency effects approached by multifractal techniques \cite{LayekS}. As quoted in \cite{LayekS}, Onsager related the Kolmogorov's -5/3 scaling law of turbulence to the fact that velocity is Hölder continuous of exponent one third. %\cite{EyinkS}.
In addition, it is well known that the isothermal surfaces in a medium of fluid turbulence are fractal above an inner scale which depends on the
Reynolds number \cite{cps91} suggesting that the same thing may happen with the fractality of the isospectral landscapes in terms of some scale depending on
the deformation parameter.

%6666666666666666666666666666666666666
%6666666666666666666666666666666666666
%6666666666666666666666666666666666666

\section{Methodology}
\subsection{Isospectral texture of energy landscapes}

We consider the very simple periodic two-dimensional Schr\"odinger equation of separable Cartesian variables \cite{Leon}
%...............1
\begin{equation}\label{separacion}
\sum_{i=1}^2\left[-\nabla^2_i+V_{-}(x_i)\right]\psi(x_i)=0,~~\nabla_i=\frac{\partial}{\partial x_i},
\end{equation}
where $V_{-}(x_i)=\cos^2(x_i)+\sin(x_i)$ is the periodic potential function and $\psi(x_i)$ are stationary solutions.
Equation (\ref{separacion}) can be factorized by the operators
%...............2
\begin{equation}
\hat{a}_i^{\dag}=-\nabla_i+\cos(x_i),~~\hat{a}_i=\nabla_i+\cos(x_i)~.
\label{prop}
\end{equation}

Next, we use the isospectral condition \cite{Mielnik}
%...............3
\begin{equation}
\label{isoesp}
\hat{a}\hat{a}^{\dag}=\hat{b}\hat{b}^{\dag}
\end{equation}
to obtain the two Riccati equations
%................4
\begin{equation}
%\cos^2(x_i)-\sin(x_i)
V_{+}(x_i)=\beta_{i}^2(x_i)+\beta_{i}'(x_i),\quad V_{+}(x_i)=\cos^2(x_i)-\sin(x_i)
\label{Riccatis}
\end{equation}
whose general solution is
%...............5
\begin{equation}
\beta_{i,g}(x_{i})=
\cos(x_i)+\phi_{i}(x_{i})=
\cos(x_i)+\frac{e^{-2\sin(x_i)}}{\gamma_i+\int_{0}^{x_i}{e^{-2\sin(x'_i)}dx'_i}}
\end{equation}
from which one can generate the deformed potential
%.........................6
\begin{equation}\label{pi}
V_{\gamma_1,\gamma_2}(x_1,x_2)=
\sum_{i=1}^2\left[V_{+}(x_i)-2\nabla_i \beta_{i,g}(x_{i})\right]~
=\sum_{i=1}^2\left[V_{-}(x_i)-2\nabla_i\phi_{i}(x_{i})\right]~.
\end{equation}
From the last two equations, one can see that there is a value of $\gamma$ where the isospectral potential becomes singular; this value is given by $\gamma_s=-2\pi I_0(2)$ \cite{Gradshteyn}, where $I_0$ is the modified Bessel function of zero order.

\begin{figure}[htb]
\centering
\subfigure[]{\includegraphics[scale=0.41]{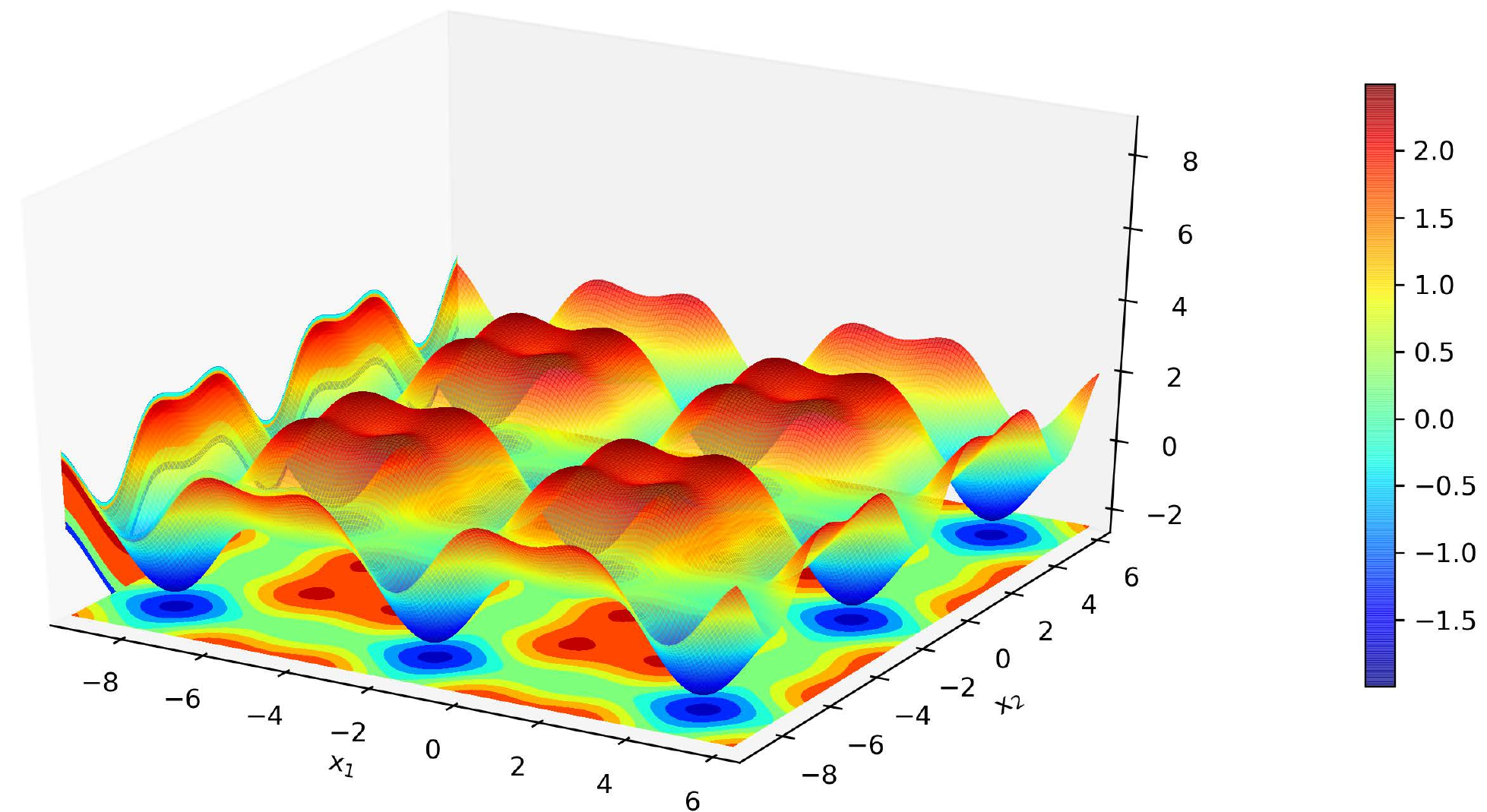}}
\subfigure[]{\includegraphics[scale=0.41]{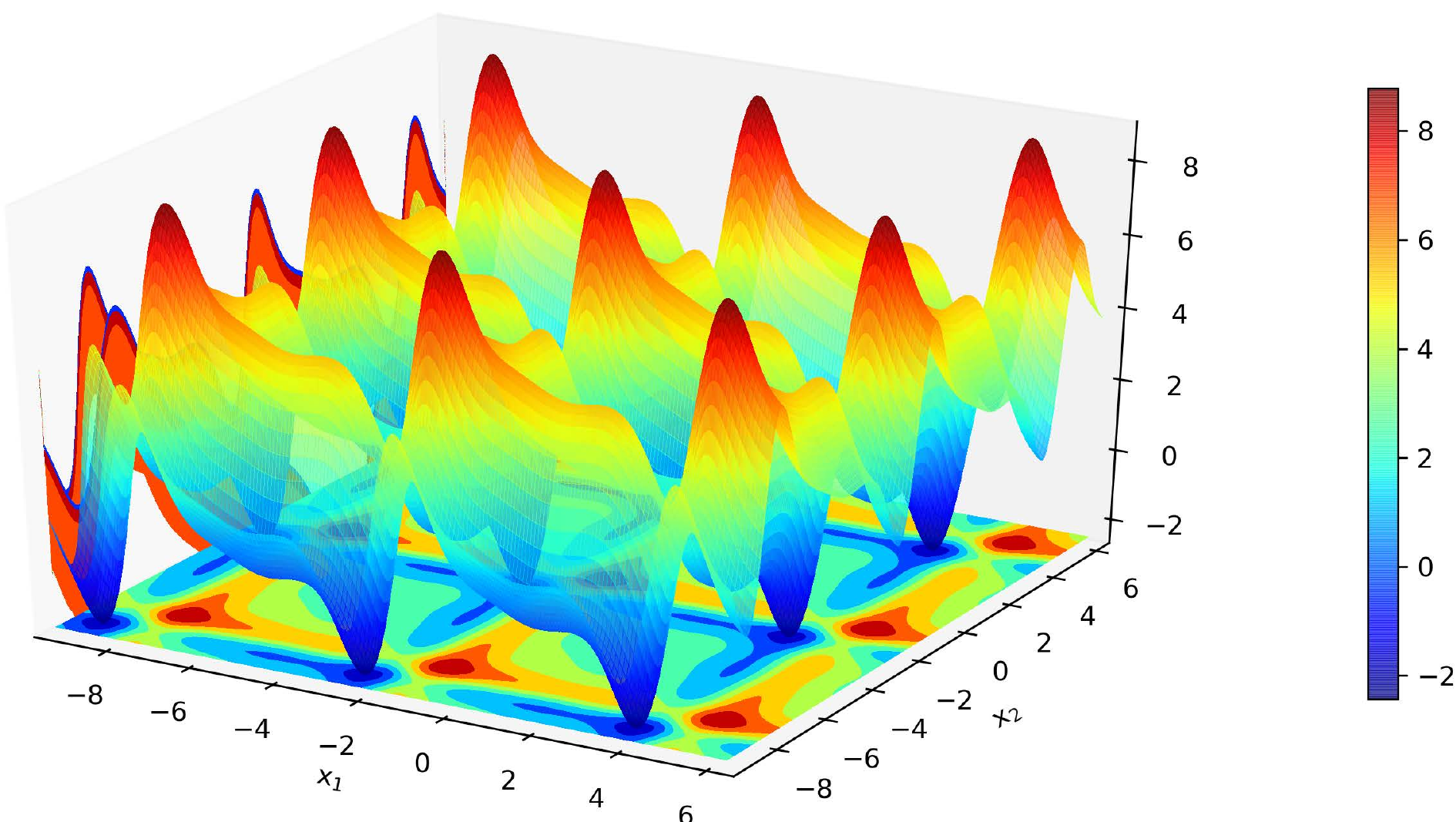}}
\caption{Three-dimensional plots of (a) the original potential $V_-(x_{1,2})$ and (b) the parametric isospectral potential $V_{\gamma_{1,2}=5}(x_{1,2}).$}
\label{vproj}
\end{figure}

The figures \ref{vproj}~(a,b) show the original potential $V_{-}(x_i)$ and the deformed one~\eqref{pi}, respectively. The contour plots have been used to classify the symmetry groups related to these potentials.

%6666666666666666666666666666666666666
%6666666666666666666666666666666666666
%6666666666666666666666666666666666666

\subsection{Hölder regularity and multifractal analysis}

Functions displaying singularities are characterized by the singularity spectrum which measure their strength through their Hölder exponents $\alpha$.
The calculation of the singularity spectrum is performed only for one coordinate since for the other one it is identical.

It is known \cite{Mallat1} that a function $f$ has a Hölder exponent
  $\alpha$ over $(c+\epsilon,d-\epsilon)$, if and only if for any $\epsilon > 0$ there exists a constant $A>0$ such that
  for $x_0 \in (c+\epsilon,d-\epsilon)$ and for $a>0$ we have $\displaystyle |W_{a,x_0}f(x)| \leq A a ^{\alpha + 1/2}$, where $W_{a,x_0}$ is
  the wavelet transform of $f$, an integral transform with wavelets as integral nuclei \cite{Mallat1}.
  In our case, with $f(x_i) = V_{\gamma_i}(x_i)$, we have  $\displaystyle |W_{a,x_0}V_{\gamma_i}(x_i)| \leq A a ^{\alpha + 1/2}$,
  which is equivalent to
%.................7
  \begin{equation}
\log|W_{a,x_0}V_{\gamma_i}(x_{i})|\leq\log A +\left(\alpha+\frac{1}{2}\right)\log(a).
  \label{wa}
  \end{equation}

  The value of $\alpha$ in \eqref{wa} corresponds to the Hölder exponent of $V_{\gamma_i}(x_{i})$ at the point $x_i=x_0$ \cite{Mallat1}.

We will also use the fractal dimension of a function $f$ defined as
%...............8
\begin{equation}
D(\alpha)=-\frac{\log N_{\alpha}(s)}{\log s}~,
\end{equation}
where $N_{\alpha}(s)$ is the number of intervals of size $s$ intersecting the set $S_{\alpha}$ of points where the function is H\"older continuous, i.e.,
its modulus of continuity is bounded by $|f(x_0) - f(x_0+s)| \leq C |s|^\alpha$ for any point $x_0\in S_{\alpha}$.

In this work, the Hölder exponents depend on the deformation parameters $\gamma_i$. Therefore we obtain a family of singularity spectra related
to the set of isospectral potentials.

%-------------------------------------------------------------------------------------------------------------------------
%-------------------------------------------------------------------------------------------------------------------------%-------------------------------------------------------------------------------------------------------------------------
%-------------------------------------------------------------------------------------------------------------------------
%-------------------------------------------------------------------------------------------------------------------------%-------------------------------------------------------------------------------------------------------------------------%-------------------------------------------------------------------------------------------------------------------------

\section{Results and discussion}
\begin{figure}[htb]
\centering
\subfigure[$Original \, potential$]{\includegraphics[scale=0.11]{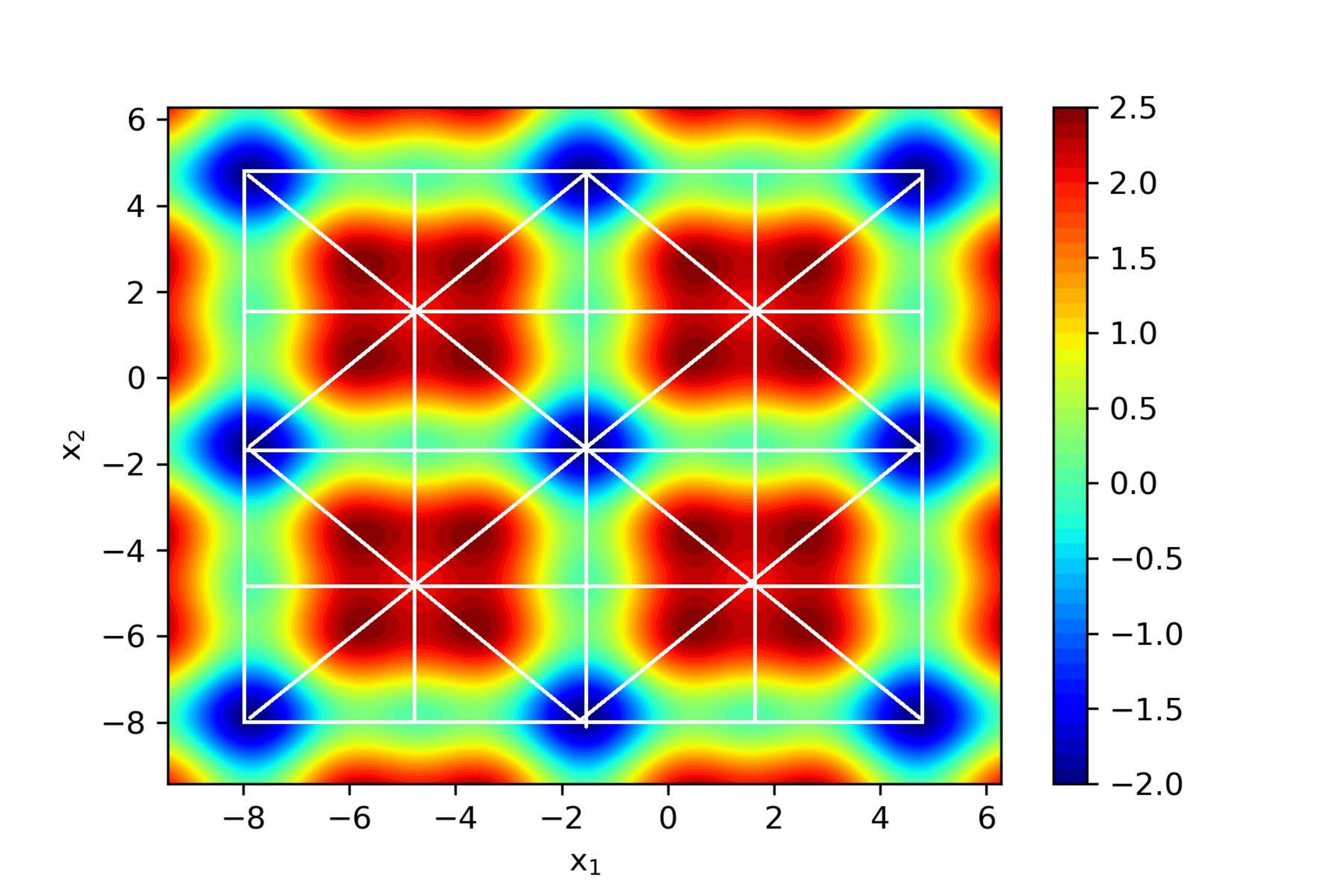}}
\subfigure[$V_{\gamma_{1,2}=\gamma_s+5}(x_{1,2})$]{\includegraphics[scale=0.11]{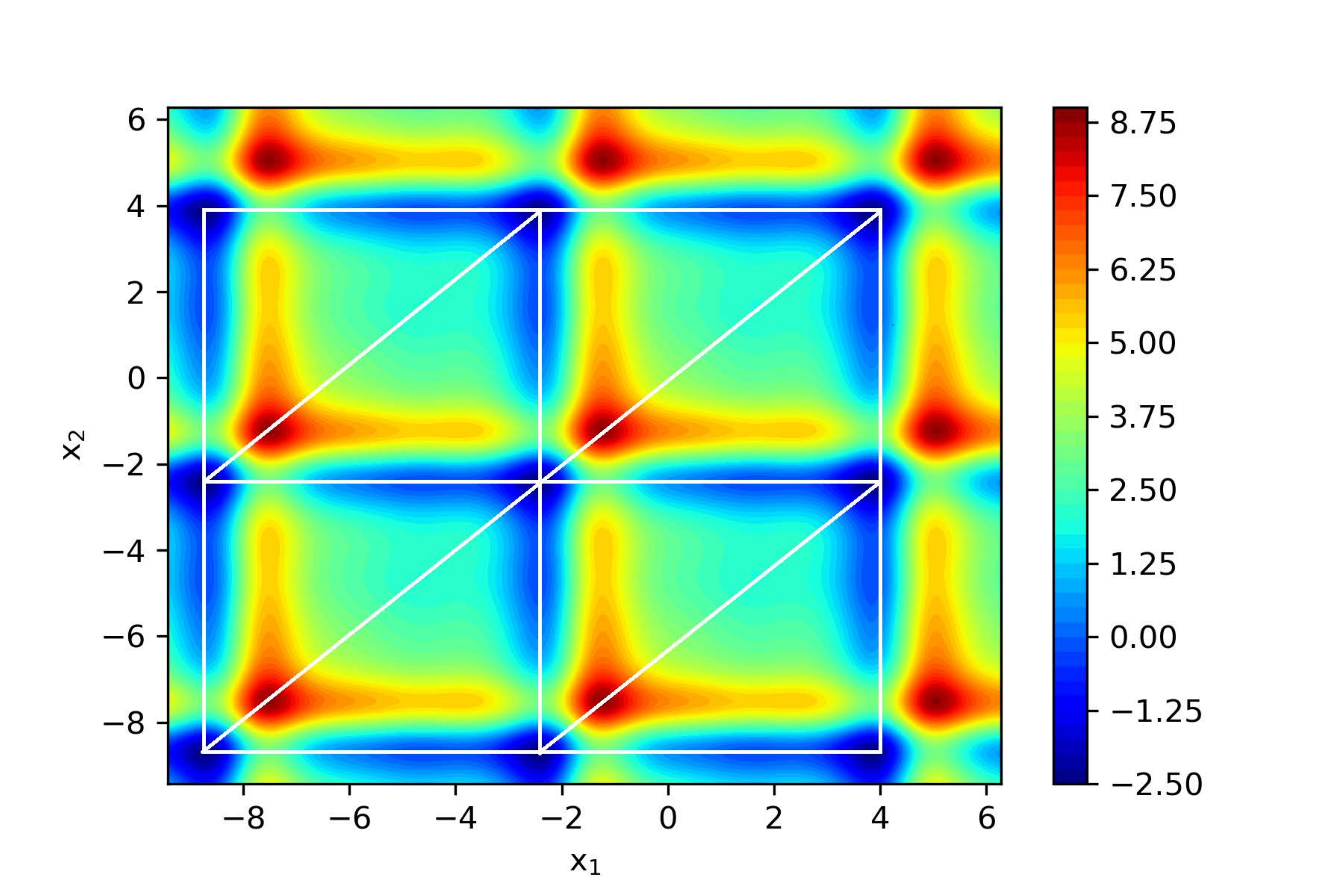}}
\subfigure[$V_{\gamma_{1,2}=\gamma_s+10}(x_{1,2})$]
{\includegraphics[scale=0.11]{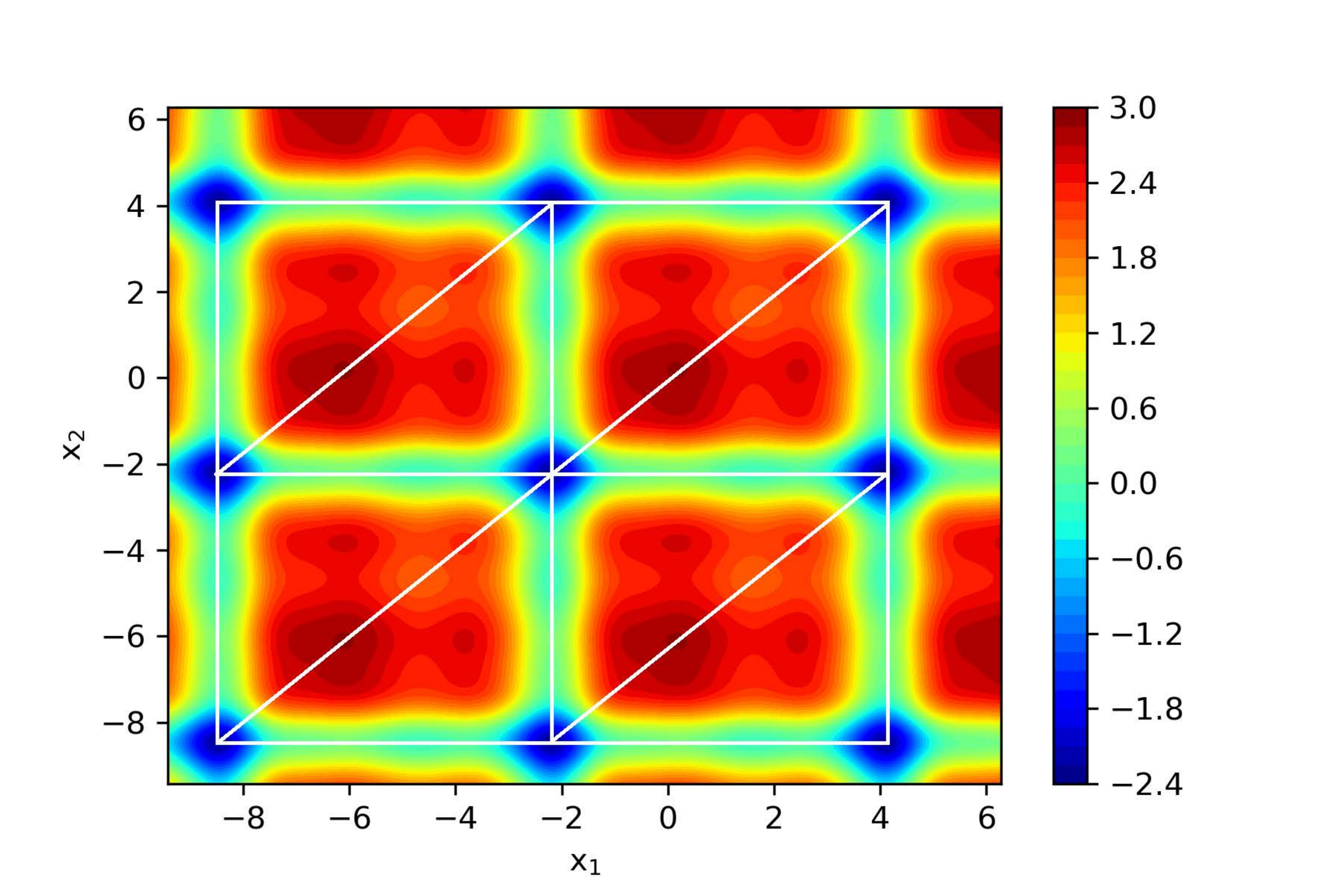}}\\
\subfigure[$V_{\gamma_{1,2}=\gamma_s+50}(x_{1,2})$]
{\includegraphics[scale=0.11]{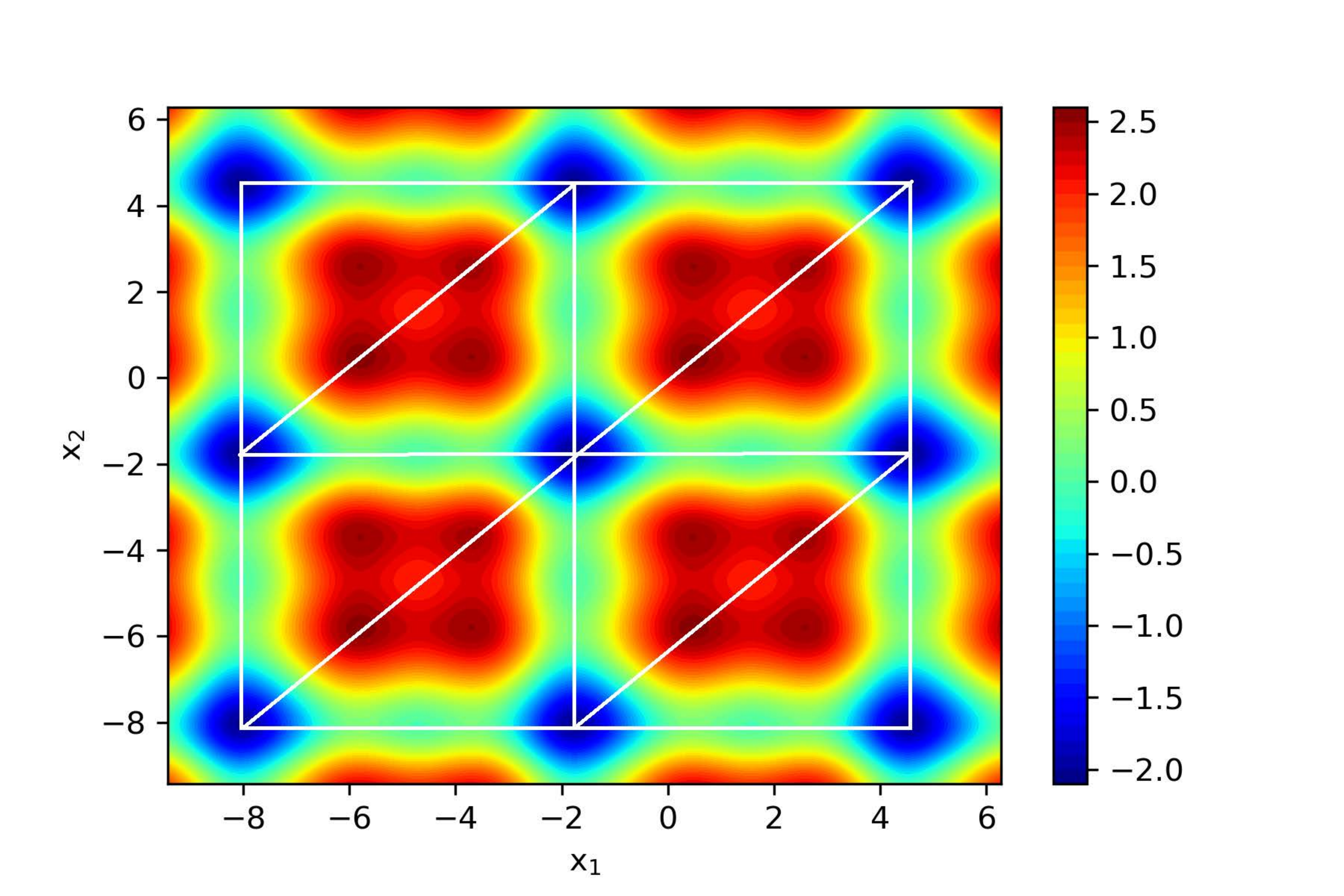}}
\subfigure[$V_{\gamma_{1,2}=\gamma_s+100}(x_{1,2})$]
{\includegraphics[scale=0.11]{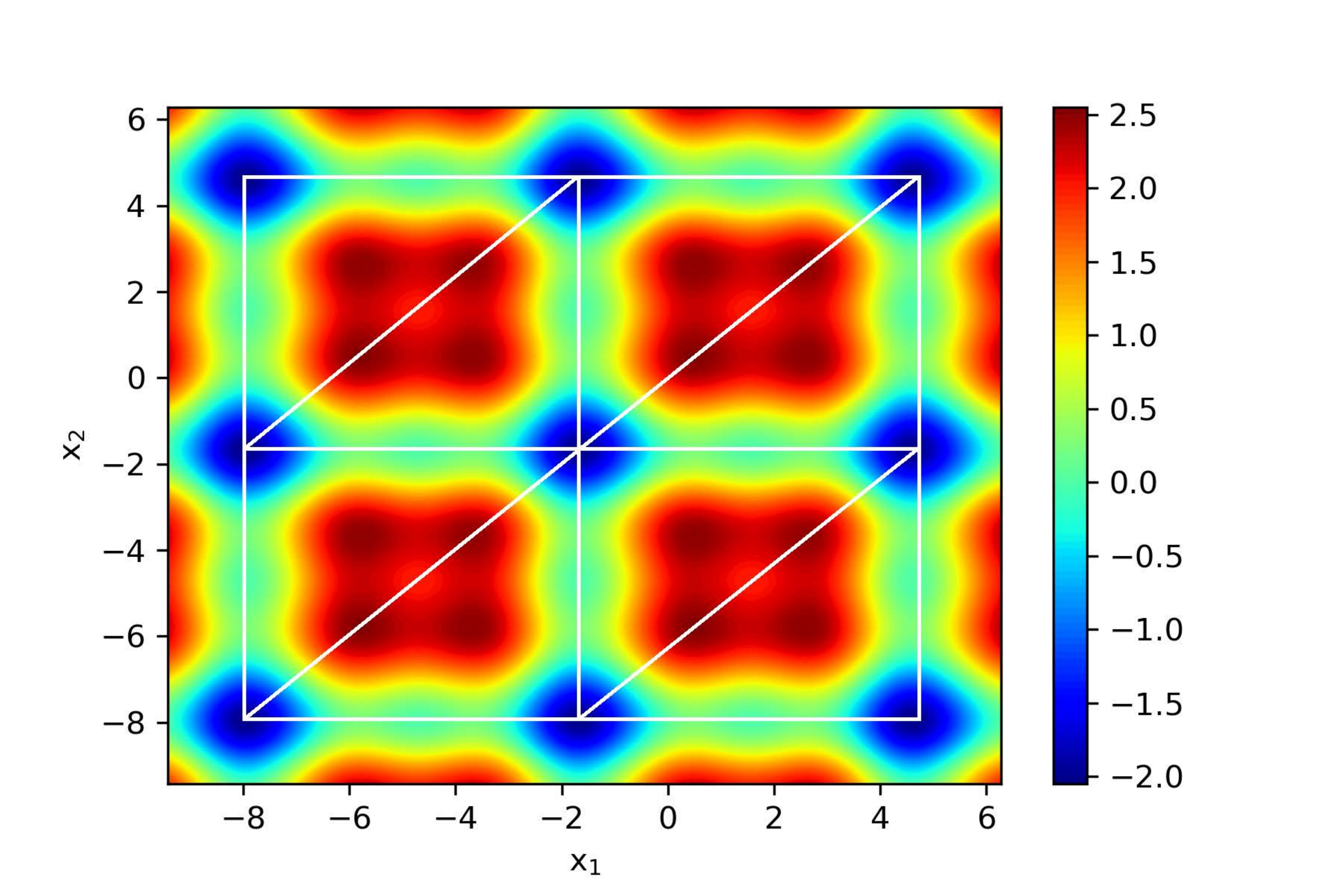}}
\subfigure[$V_{\gamma_{1,2}=\gamma_s+150}(x_{1,2})$]
{\includegraphics[scale=0.11]{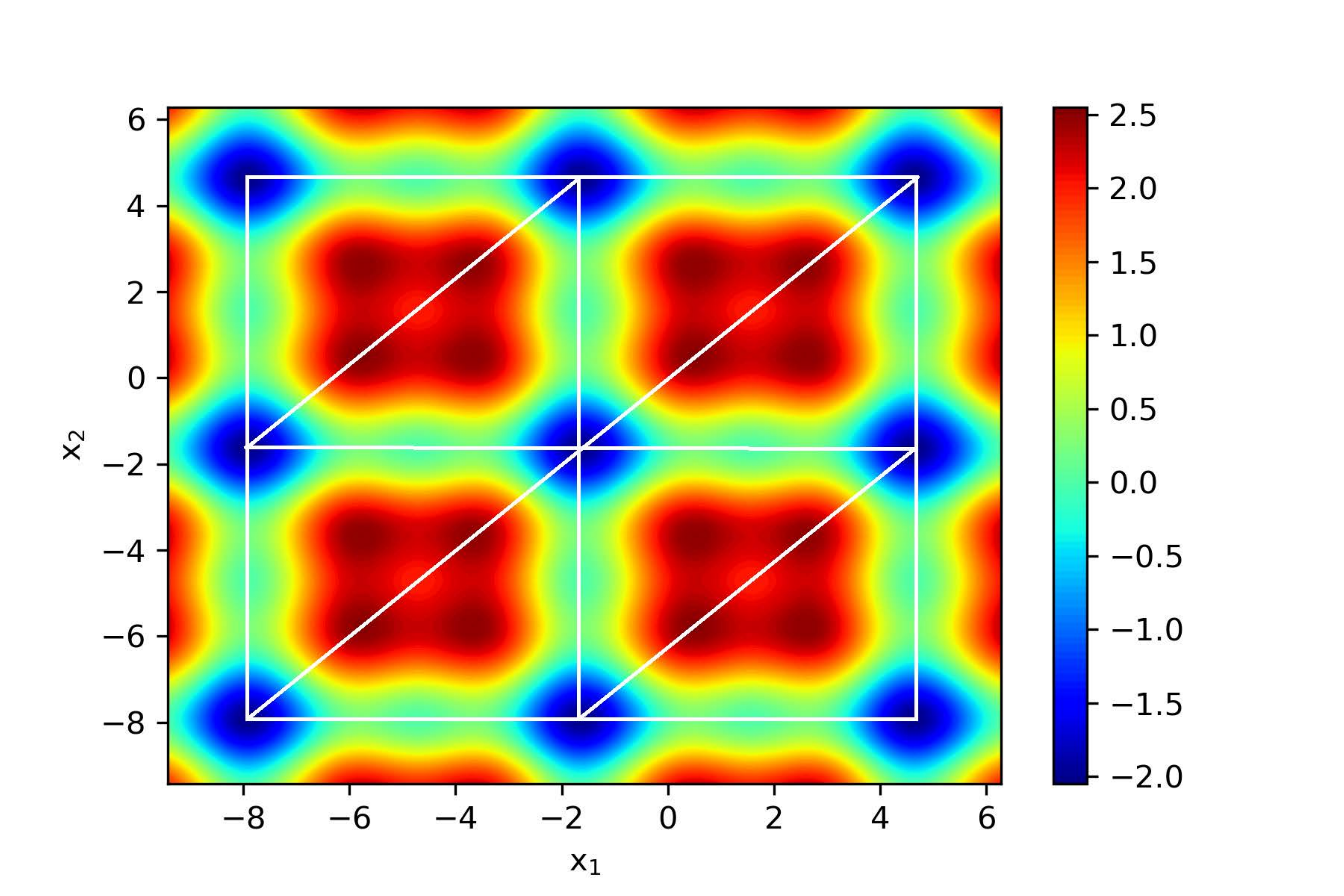}}
\caption{Convergence to the original tile.}
\label{ev1}
\end{figure}

The plots in figure~\ref{ev1} (a-h) represent the evolution of the symmetry on the tiles as a function of the deformation parameter $\gamma$, starting with the tiles of symmetry group $P_{4mm}$ related to the original potential. Because of the isospectral transformation, the group is modified to $P_m$ symmetry group, but when the deformation parameter $\gamma$ takes values bigger than a critical value a return to the original group $P_{4mm}$ can be seen in figure~\ref{ev1}.
The symmetry group classification is according to \cite{symm}.
Since the critical value of $\gamma$ cannot be determined only from the contour plots of the parametric potentials, we analyze their Hölder regularity to find a very accurate value.

\begin{figure}[htb]
\subfigure[]
{\includegraphics[scale=0.58]{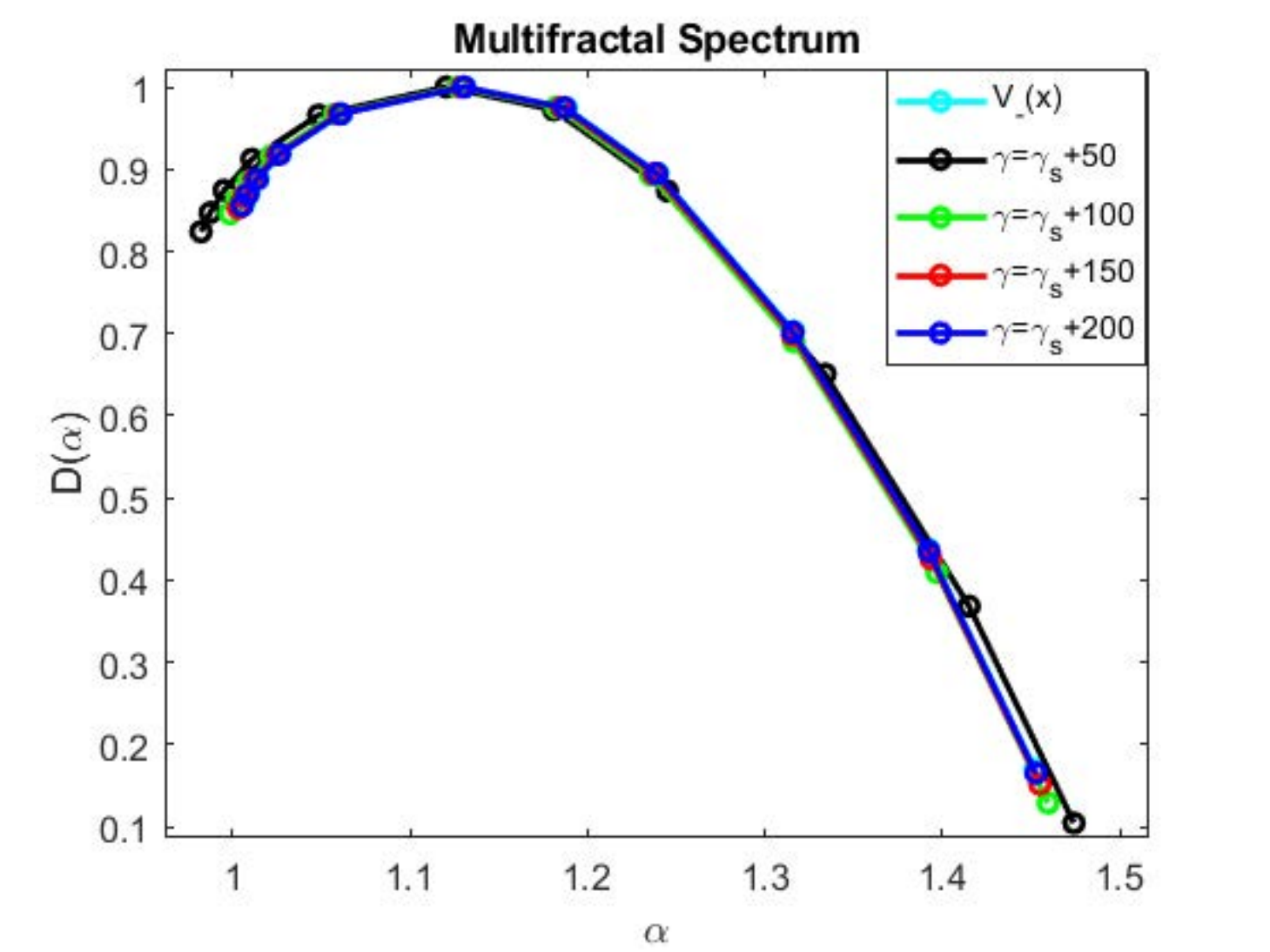}}
\subfigure[]
{\includegraphics[scale=0.58]{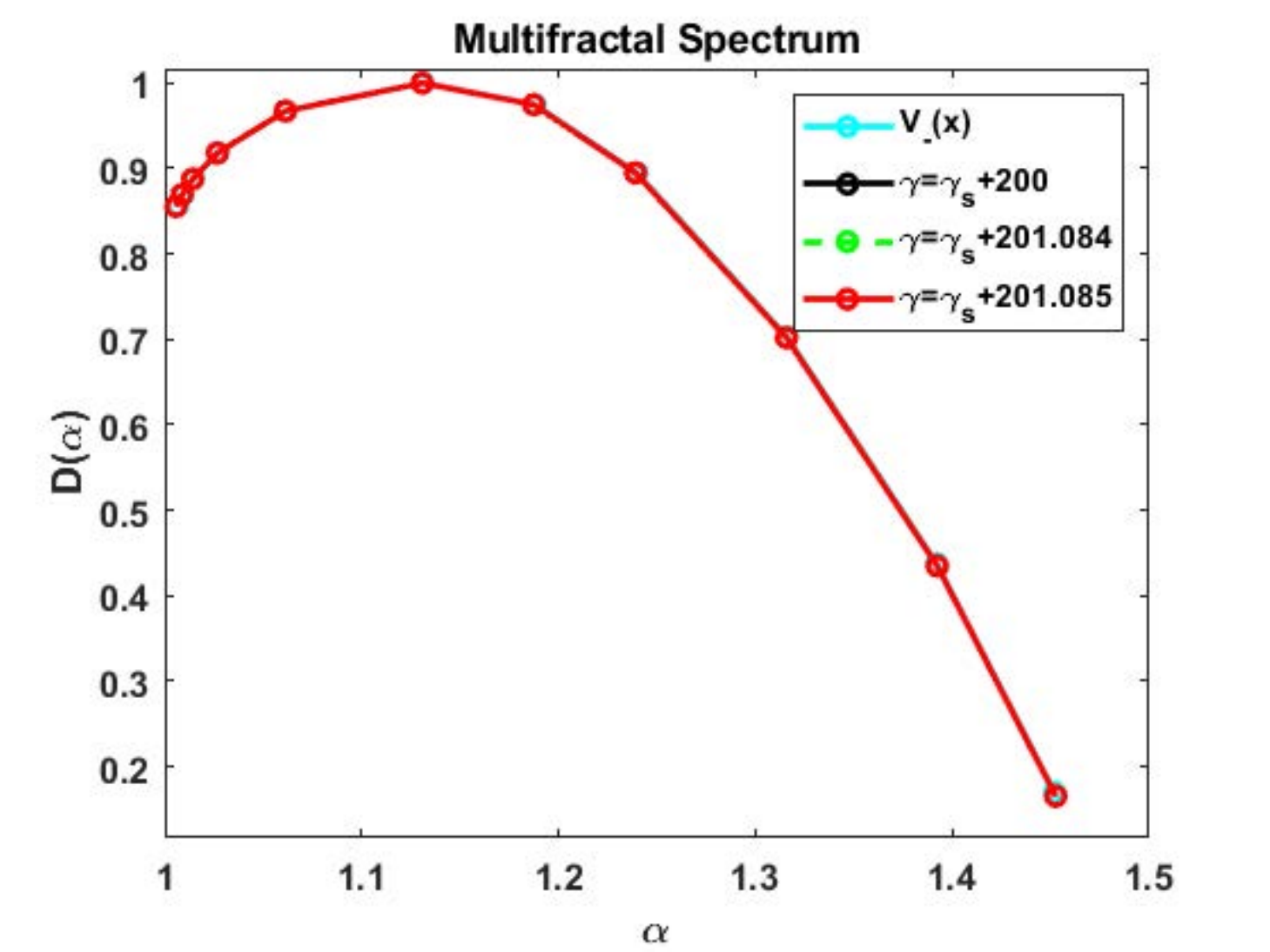}}
\caption{Convergence to the multifractal spectrum of the original potential $V_-(x)$.}
\label{mfctl}
\end{figure}

\begin{figure}[htb]
\subfigure[]
{\includegraphics[scale=0.58]{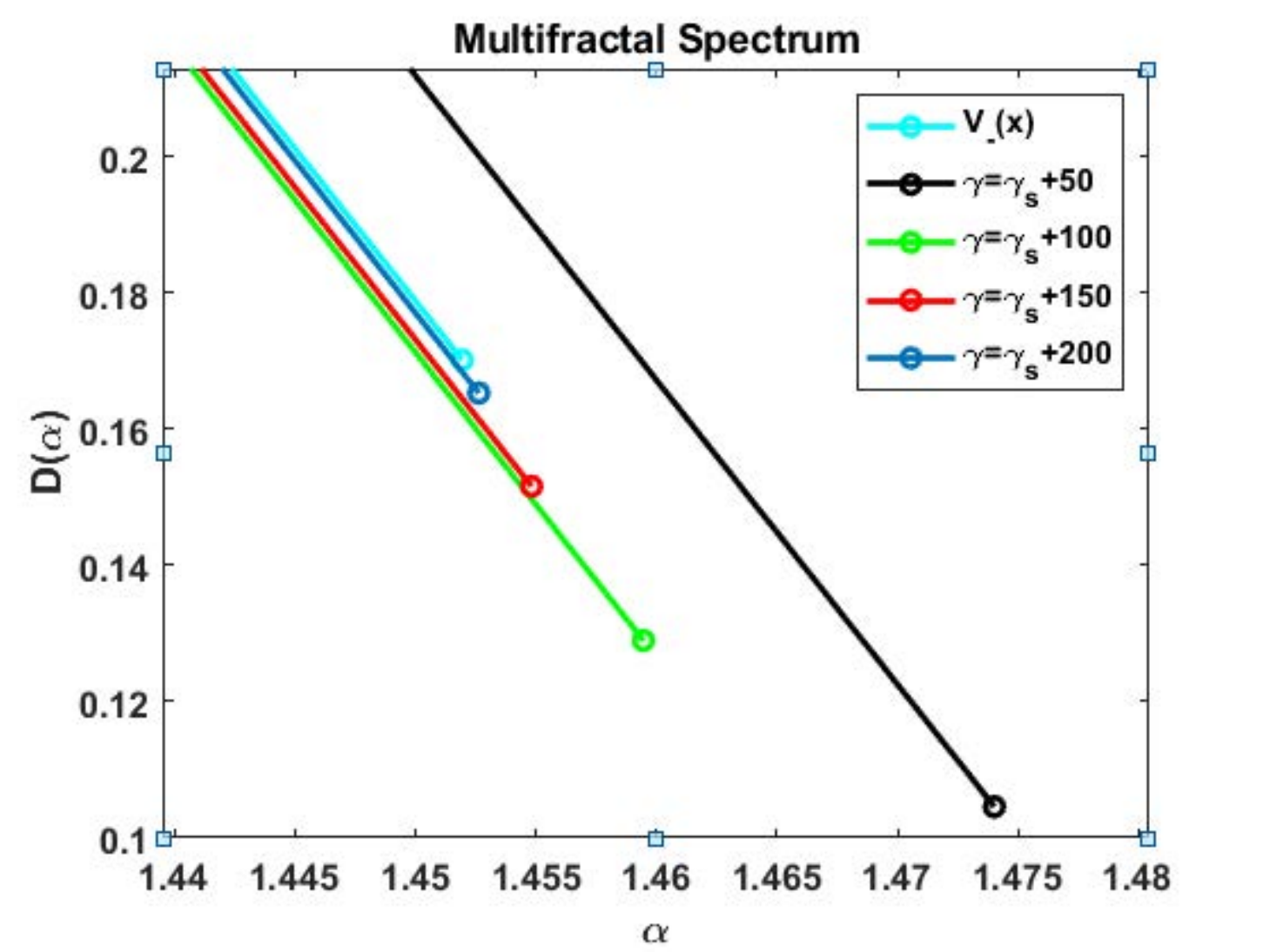}}
\subfigure[]
{\includegraphics[scale=0.58]{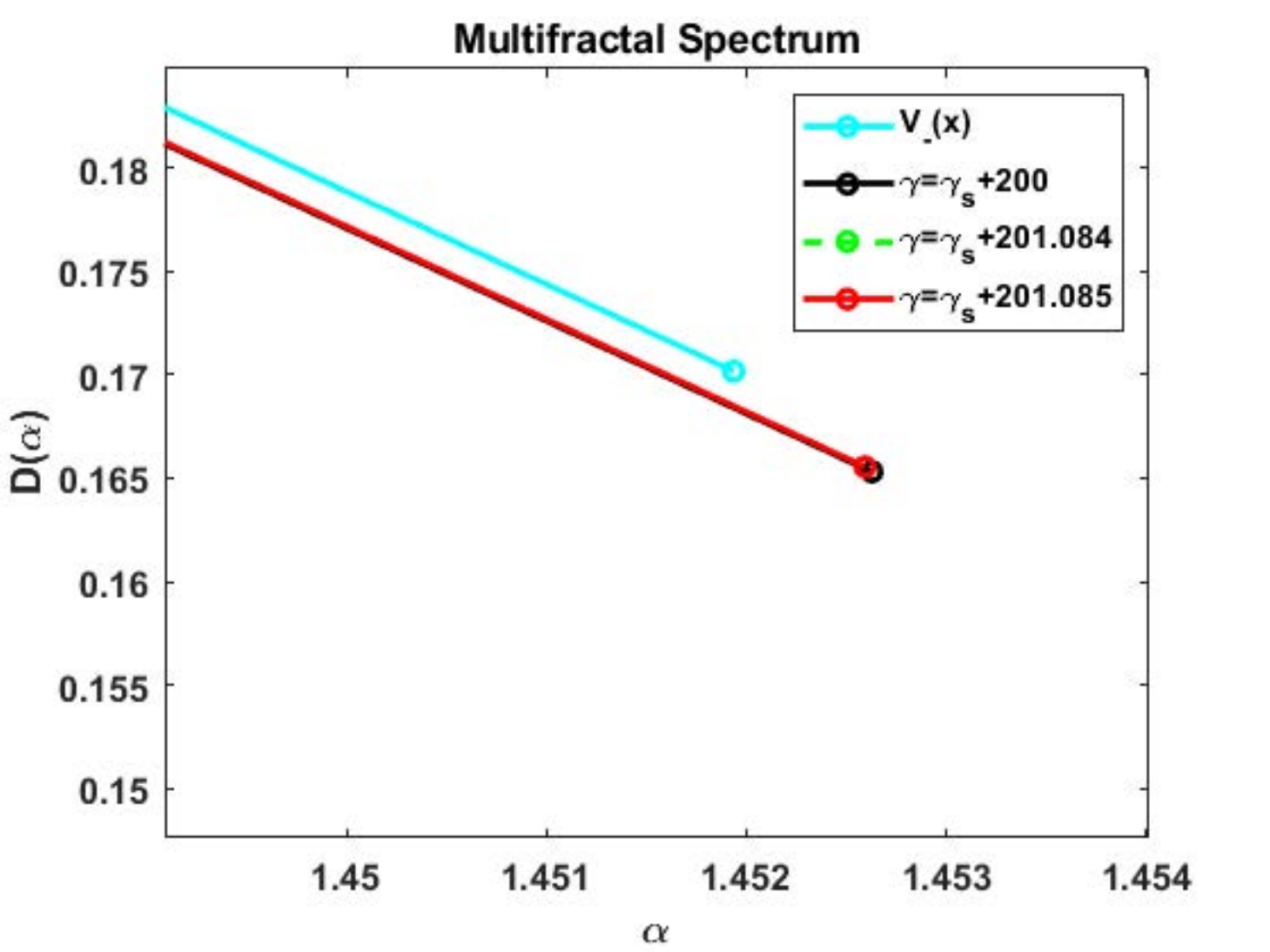}}
\caption{Amplification of the convergence region for the multifractal spectrum.}
\label{mfctlzoom}
\end{figure}

The original potential has an associated multifractal spectrum given in figure~\ref{mfctl} that is used to find the approximated critical value of the deformation parameter. The precision of this method can be seen in the figures~\ref{mfctlzoom} that show the region where the multifractal spectra are different for the studied potentials.

The results show that the multifractal spectrum of the original potential is recovered when the deformation parameter takes rather big values. To establish a precise value for the deformation parameter in terms of local regularity we use the method of lines of modulus maxima with the Mexican hat wavelet. The method provides a full characterization of the potential generating a table of Hölder exponent values for the considered potential whose positions vary as a function of the deformation parameter. Moreover, we can introduce a condition of equivalence of two potentials as follows:

\textit{Two potentials are equivalent when the number of singularities and Hölder exponents of the potentials are the same and are located in the same neighborhoods.}

We thus present in Table 1 the Hölder exponent values and their locations for the original potential and the isospectral parametric potential at the determined values of the deformation parameter at which the values of Hölder exponents ($\alpha$) are as close as numerically possible to the values
of the original potential, while their position ($x_0$) may be considered in a small neighborhood of the original locations.
Proceeding in this way, we find that the critical value of $\gamma$ at which the symmetry group of the original potential is recovered is in the interval %between
$\gamma_s+201.084<\gamma<\gamma_s+201.085$. %\eqref{tab:tabla}.


%\begin{landscape}
\begin{table}[htb]
\centering
\small{
\begin{tabular}{ |c|c|c|c|c|c|c|c|c|c|c|c|c|c|c|c|c|c|c|c| }
  \hline
  \multicolumn{2}{|c|}{$V_-(x)$}&
  \multicolumn{2}{|c|}{$\gamma_s+10$}&
  \multicolumn{2}{|c|}{$\gamma_s+50$}&
  \multicolumn{2}{|c|}{$\gamma_s+100$}&
  \multicolumn{2}{|c|}{$\gamma_s+201.084$}&
  \multicolumn{2}{|c|}{$\gamma_s+201.085$}
  \\
  \hline
$x_0$   &$\alpha$&$x_0$&$\alpha$&$x_0$ &$\alpha$& $x_0$   &$\alpha$&  $x_0$  &$\alpha$& $x_0$ &$\alpha$\\
    \hline
  5&1.978&32  &1.575 &5   &1.909  & 5   & 1.955  & 5   & 1.975  & 5&1.975     \\
170&1.963&118 &1.844 &172 &1.959  & 171 & 1.959  & 170 & 1.963  & 170&1.963   \\
314&1.960&193 &1.958 &313 &1.960  & 314 & 1.960  & 314 & 1.960  & 314&1.960   \\
459&1.966&310 &1.956 &460 &1.964  & 460 & 1.965  & 459 & 1.966  & 459&1.966   \\
619&1.999&459 &1.954 &599 &1.924  & 609 & 1.996  & 617 & 1.998  & 617&1.999   \\
%   &     &573 &1.656 & &  &  & && &     &       \\
  \hline
  \multicolumn{12}{|c|}{For $\gamma_s+201.085$, all $\alpha$'s but one are identical to those of $V_-(x)$ up to three decimal digits.}
  \\
    \hline
  \end{tabular}
\caption{Hölder exponents ($\alpha$) and their position ($x_0$) for the original potential and five members of the family of parametric isospectral potentials, of which the last two are separated by only $1\cdot 10^{-3}$ in the deformation parameter.}
\label{tab:tabla}
}
\end{table}
%\end{landscape}

%6666666666666666666666666666666666666
%6666666666666666666666666666666666666

\section{Conclusion}

We have shown through a simple illustrative example that the contour plots of parametric potentials which are supersymmetric isospectral to periodic potentials provide interesting energy landscapes in which the electrons may move in a material planar slab. They display symmetry properties of much interest in materials science such as symmetry breaking and the recovery of the symmetry group related to the initial landscape.

As shown here, the multifractal analysis provides a high-precision measure of the convergence of the deformed potentials to the original one when the deformation parameter takes big values. Using this kind of analysis, we can establish a relationship between the multifractal spectrum and the symmetry breaking in terms of equivalence of their multifractal spectra. Highly precise values of the critical deformation parameter can be obtained through the convergence of Hölder exponents for the deformed potential to those of the original one.

In materials science, there are other issues that can be approached by multifractal analysis. An example of transition of symmetry groups $P_m \to P_{4mm}$ as in our study occurs in a phase transition of ferroelectric type due to temperature changes in a ceramic solid solution \cite{Ge}.

Furthermore, the parametric isospectral (supersymmetric) energy landscapes that we have introduced here have potential applications to areas such as electron transport  in semiconductor superlattices, optical superlattices, and trapped ultracold atoms, to name just a few.
Besides, more complicated isospectral energy landscapes can be constructed by using factorization operators with more periodic components.

\subsection*{Credit authorship contribution statement}

\noindent {\bf J. de la Cruz}: Writing of initial version, Methodology, Calculations.

\noindent {\bf J.S. Murgu\'{\i}a}: Supervision, Validation, Calculations.

\noindent {\bf H.C. Rosu}: Writing of the final draft, Supervision, Formal analysis.

\subsection*{Declaration of competing interests}

The authors declare they have no known competing financial interests or personal relationships that could
have appeared to influence the work reported in this paper.

\subsection*{Aknowledgements}

The first author acknowledges the financial support of CONACyT through a doctoral fellowship.
This paper was partially supported by CONACyT funds from grant CB 2017–2018 A1-S-45697.
The authors thank the referees for important remarks that helped to refine this work.


\begin{thebibliography}{4}
\bibitem{Mielnik}
B.~Mielnik, Factorization method and new potentials with the oscillator spectrum, J. Math. Phys. 25 (1984) 3387-3389.
\bibitem{rmc}
H.C.~Rosu, S.C.~Mancas, P. Chen, One-parameter families of supersymmetric isospectral potentials from Riccati solutions in function composition form,
Ann. Phys. 343 (2014) 87-102.
\bibitem{cz14} T.L.~Curtright and C.K.~Zachos, Branched Hamiltonians and supersymmetry, J. Phys. A: Math. Theor. 47 (2014) 145201.
\bibitem{Sako-Suzuki} A.~Sako and T.~Suzuki, Recovery of full $N=1$ supersymmetry in non(anti-)commutative superspace,
JHEP 11 (2004) 010.
\bibitem{Bevilaqua} L.~Ibiapina Bevilaqua, A.C. Lehum, A.J.~da Silva, Soft supersymmetry breaking in the nonlinear sigma model,
Phys. Lett. B 789 (2019) 150-153.
\bibitem{He}
J.S. Murguía and J. Urías, On the wavelet formalism for multifractal analysis, Chaos 11 (2001) 858-863.
\bibitem{LayekS}
G.C. Layek and Sunita, Multifractal cascade symmetry model for fully developed turbulence, Fractals 26 (2018) 1850070.
\bibitem{cps91}
P. Constantin, I. Procaccia, K.R. Sreenivasan, Fractal geometry of isoscalar surfaces in turbulence: Theory and experiments,
Phys. Rev. Lett. 67 (1991) 1739-1742.
\bibitem{Leon}
J.~Socorro, M.A. Reyes, C.V. Mora, E. Condori, Supersymmetric quantum mechanics: two factorization schemes and quasi-exactly solvable potentials, in Panorama of contemporary quantum mechanics - concepts and applications, editor T.T. Truong, IntechOpen, 2019.
\bibitem{Gradshteyn}
I.S.~Gradshteyn, I.M. Ryzhik, Table of integrals, series, and products, p. 496, formula 3.~937.~2, Academic Press, 2007.
\bibitem{Mallat1}
S.~Mallat, A wavelet tour of signal processing, chapter 6, third edition, Academic Press, 2009.
\bibitem{symm}
D.~Schattschneider, The plane symmetry groups: Their recognition and notation, Amer. Math. Monthly 85 (1978) 439-50.
\bibitem{Ge}
W.~Ge, Y. Ren, J. Zhang, C.P. Devreugd, J. Li, D. Viehland, A monoclinic-tetragonal ferroelectric phase transition in lead-free
($K_{0.5}Na_{0.5})NbO_3-x$\%$LiNbO_3$ solid solution, J. Appl. Phys. 111 (2012) 103503.

\end{thebibliography}
\end{document}